\documentclass[12pt]{spieman}  
\usepackage{amsmath,amsfonts,amssymb}
\usepackage{graphicx}
\usepackage{setspace}
\usepackage{tocloft}

\title{Baseline Requirements For Detecting Biosignatures with the HabEx and LUVOIR Mission Concepts}

\author[a,*]{Ji Wang}
\author[a]{Dimitri Mawet}
\author[b,c]{Renyu Hu}
\author[a]{Garreth Ruane}
\author[a]{Jacques-Robert Delorme}
\author[a]{Nikita Klimovich}

\affil[a]{Department of Astronomy,  California Institute of Technology, Pasadena, CA 91125, USA}
\affil[b]{Jet Propulsion Laboratory, California Institute of Technology, Pasadena, CA 91109, USA}
\affil[c]{Division of Geological and Planetary Sciences, California Institute of Technology, Pasadena, CA 91125, USA}

\cftpagenumbersoff{figure}
\cftpagenumbersoff{table} 
\begin{document} 
\maketitle

\begin{abstract}
A milestone in understanding life in the universe is the detection of biosignature gases in the atmospheres of habitable exoplanets. Future mission concepts under study by the 2020 decadal survey, e.g., HabEx and LUVOIR, have the potential of achieving this goal. We investigate the baseline requirements for detecting four molecular species, H$_2$O, O$_2$, CH$_4$, and CO$_2$, assuming concentrations of these species equal to that of modern Earth. These molecules are highly relevant to habitability and life on Earth and other planets. Through numerical simulations, we find the minimum requirements of spectral resolution, starlight suppression, and exposure time for detecting biosignature and habitability marker gases. The results are highly dependent on cloud conditions. A low-cloud case is more favorable because of deeper and denser lines whereas a no-cloud case is the pessimistic case for its low albedo. The minimum exposure time for detecting a certain molecule species can vary by a large factor ($\sim$10) between the low-cloud case and the no-cloud case. For all cases, we provide baseline requirements for HabEx and LUVOIR. The impact of exo-zodiacal contamination and thermal background is also discussed and will be included in future studies.   
\end{abstract}


{\noindent \footnotesize\textbf{*}Further author information: e-mail:\linkable{ji.wang@caltech.edu} }

\begin{spacing}{2}   

\section{Introduction}
\label{sect:intro}  %

Thousands of exoplanets have been discovered to date and many more will be detected by future missions. The focus of exoplanet studies are shifting towards understanding the statistical properties of exoplanets as a population and detailed characterization for scientifically-compelling nearby systems. One of the primary goals of the latter is to study the chemical composition of exoplanet atmospheres and the implications for habitability and life.

Habitability requires a surface temperature that is suitable for life and the existence of liquid water (H$_2$O). Therefore, H$_2$O is a high priority species to identity with future telescopes. In addition, biosignature gases, such as oxygen (O$_2$) and methane (CH$_4$), are highly indicative of life when co-existing out of thermodynamic equilibrium\cite{DesMarais2002}. 
Carbon dioxide (CO$_2$) is the most prominent greenhouse gas and is a key species to identify in habitable exoplanets. The total concentration of CO$_2$ is expected to differ on a habitable planet based on its position in the habitable zone. A low CO$_2$ concentration is expected for planets near the inner edge of the habitable zone and a high CO$_2$ concentration is expected for planets near the outer edge\cite{Kopparapu2013}. However, a high concentration of CO$_2$, together with high stellar UV flux, may produce abiotic O$_2$\cite{Hu2012b,DomagalGoldman2014, Tian2014, Harman2015}. In this case, quantifying CO$_2$, O$_2$, and other gases will help rule out false positive scenarios.
Future missions to detect and characterize habitable planets will have the ability to identify biosignature gases. However, there are several outstanding technical challenges:
\begin{itemize}
    \item {\textbf{Bigger aperture}} Aperture size of future space missions will be marginally adequate to identify certain biosignature gases. For example, as shown in this paper, it is extremely challenging to detect CH$_4$ in certain cases. 
    \item {\textbf{Cooler system}} Despite much stronger biosignature signals in the near infrared, it is likely that future large aperture space missions (4m-6.5m HabEx and 8m-16m LUVOIR) will not have the capability to operate at wavelengths longer than 1.8 $\mu$m because of increasing thermal background from the telescope system. To operate beyond 1.8 $\mu$m and have reasonably low thermal background, complicated cooling systems would be required , which may drastically increase the cost of the mission.  
    \item {\textbf{Higher spectral resolution
    }} Identifying biosignature gases requires spectroscopic analysis of light from exoplanets. Most proposed dispersing elements for space missions to date have relatively low spectral resolution, e.g., R $\sim$ 70 for WFIRST-IFS\cite{McElwain2016}, which does not take advantage of rich spectral lines of the molecules of interest.
    \item {\textbf{Better performance of adaptive optics systems}}
    Ground-based extremely large telescopes (ELTs) or Giant Segment Mirror Telescopes (GSMTs) will face severe challenges in breaking the current starlight suppression floor at 10$^{-5}-10^{-6}$ level, which is set by adaptive optics system temporal bandwidth to correct for the Earth's atmospheric turbulence. 
\end{itemize} 

   \begin{figure}
   \begin{center}
   \begin{tabular}{c}
   \includegraphics[height=12cm]{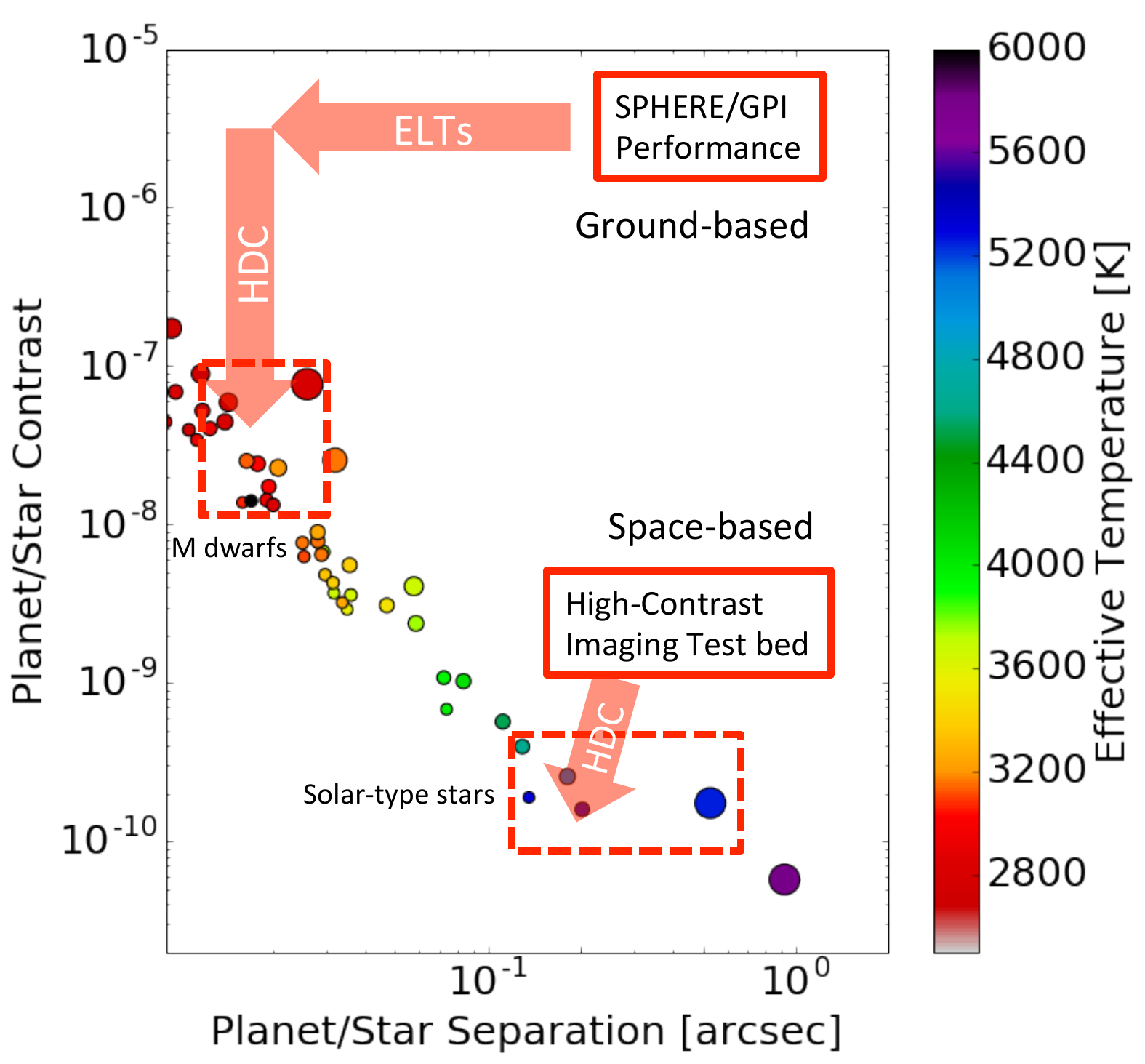}
   \end{tabular}
   \end{center}
   \caption[Planet-star contrast and angular separation for a hypothetical Earth-sized planet in the habitable zone around each star within 5 pc]
   { \label{fig:con_sep} 
   Planet-star contrast and angular separation for a hypothetical Earth-sized planet in the habitable zone around each star within 5 pc\footnote{\url{https://www.naoj.org/staff/guyon/}}. Each data point is colored by host star effective temperature and the size is scaled with distance. Red solid-lined boxes mark state-of-the-art performance for ground-based direct imaging instruments and test beds for future space-based missions. Red dashed-line boxes indicate the notional objectives for future ground-based missions (i.e., planets around M dwarfs) and for future space missions (i.e., planets around solar-type stars)}
   \end{figure} 

Fig. \ref{fig:con_sep} summarizes the state-of-the-art performance of direct imaging instruments or testbeds (solid-lined boxes) and the requirements to detect a habitable planet (dashed-lined boxes). For both ground-based and space-based missions, there is a contrast gap between state-of-the-art performance and the planet-star contrast objective. 

High dispersion coronagraphy (HDC) is a recent technical development that is designed to bridge the contrast gap\cite{Wang2017}. HDC combines high contrast imaging (HCI), a single-mode fiber injection unit (FIU) \cite{Mawet2017}, and high resolution spectroscopy (HRS) to filter out stellar light and extract the planet's signal. Specifically, HCI suppresses stellar light and spatially separates the planet from its host star. The FIU filters out stellar noise at the planet location since the electric field of a stellar speckles does not couple to the fundamental mode of a single-mode fiber, whereas up to $\sim80\%$ of the planet light couples into the fiber. HRS further distinguishes planet signal from stellar signal by its unique spectral features such as absorption lines and radial velocity. Using this three-pronged starlight suppression, HDC can achieve the high sensitivity  to study terrestrial planets in the habitable zone. 

This paper focuses on the application of HDC on future space telescopes, e.g., HabEx and LUVOIR. The fundamental question we attempt to address here is whether these missions can detect four potential biosignature and habitability marker gases: O$_2$, CH$_4$, H$_2$O, and CO$_2$, and set mission requirements to achieve this goal. The four molecular species that are chosen have a relatively high concentration in the Earth's atmosphere and are significant signs of life. In addition, these species have been detected in earthshine and spacecraft observations of our own planet\cite{Seager2016}. Detecting other biosignature gases , which have much lower concentrations, is expected to be far more challenging for exoplanets.

The paper is organized as follows. In \S \ref{sec:mol}, we briefly describe the four molecular species we investigate in this paper and the model we use to generate the spectra for our HDC simulations. In \S \ref{sec:hdc}, we briefly describe our HDC simulation approach. More details can be found in previous work on HDC\cite{Wang2016, Wang2017}. Results are given in \S \ref{sec:result} followed by summary and conclusion in \S \ref{sec:summary}.

\section{MOLECULAR SPECIES OF INTEREST}
\label{sec:mol}

\subsection{Model Description}

The spectra of Earth-like exoplanets are generated by an atmospheric chemistry and radiative transfer model~\cite{Hu2012b, Hu2013, Hu2014}. We calculate the molecular abundance as a function of altitude, controlled by photochemical and disequilibrium chemistry processes~\cite{Hu2012b}. The model atmosphere has been compared with terrestrial measurements in the mid-latitude and closely resembles the present-day Earth~\cite{Hu2012b}. We calculate the disk-integrated reflected light spectra with an eighth-order Guassian integration and $\delta$-2-stream approximation. We include the opacities of CO$_2$, O$_2$, H$_2$O, and CH$_4$ and calculate the planetary flux at a spectral resolution of $R=\lambda/\Delta\lambda=500,000$, which is high enough to resolve individual spectral lines of the aforementioned species over $\lambda=0.5$ - 5 $\mu$m.

It is well known that Earth’s disk-averaged spectrum cannot be reproduced by a single surface type or cloud deck~\cite{Turnbull2006,Tinetti2006,Robinson2011}. Rather, it requires a combination of a cloud-free surface that is poorly reflective (i.e., the ocean), a highly reflective “low cloud” at $\sim4$ km mimicking the cumulus clouds, and a “high cloud” at $\sim12$ km mimicking the cirrus clouds~\cite{Turnbull2006}. We follow this treatment and assume the three types of surface or cloud. We assign equal weights to the three components to derive a reasonable account for an ``Earth-like'' planet. While the purpose is to reproduce Earth's spectrum, the resulting spectrum is generally consistent with the Earthshine experiments~\cite{Turnbull2006,Tinetti2006} as well as EPOXI measurements~\cite{Robinson2011}.

\subsection{Earth Spectrum By Molecule}
\label{sec:EarthSpec}

\begin{figure}
   \begin{center}
   \begin{tabular}{c}
   \includegraphics[height=16cm]{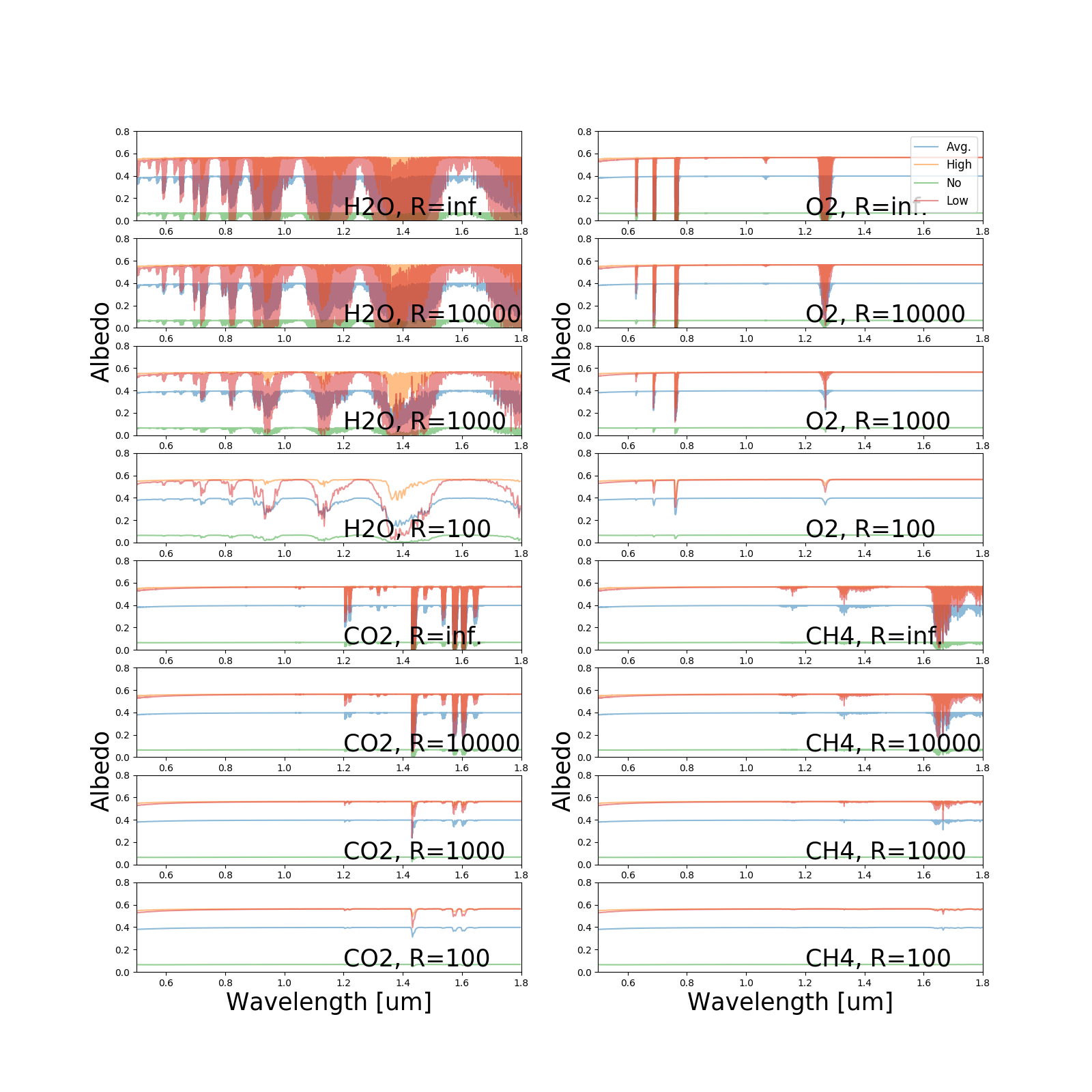}
   \end{tabular}
   \end{center}
   \caption[Albedo spectra] 
   { \label{fig:mol_res} 
   Albedo spectra for H$_2$O (top-left)
    O$_2$ (top-right), CO$_2$ (bottom-left), and CH$_4$ (bottom-right). Colors represent  cases with different cloud conditions. Each panel has 4 rows to show the influence of spectral resolution on spectral features. Top row is for R$>$500,000 followed by R of 10000, 1000, and 100, respectively.}   \end{figure}

We consider four molecular species in an Earth's atmosphere, CO$_2$, O$_2$, H$_2$O, and CH$_4$. Their spectra at different spectral resolutions are shown in Fig. \ref{fig:mol_res}. Observable line density and line depth decrease as spectral resolution decreases, making detecting certain molecules difficult at low spectral resolutions. Conversely, planet signal is dispersed to more pixels at high spectral resolutions. In such a case, molecule detectability is generally limited by detector noise. {{Since there are a number of zero- or low-noise detectors that potentially allow us to overcome the detector noise limited case\cite{Rauscher2016}, we assume noiseless detectors in the HDC simulations in this paper.}} 


Planet reflection spectrum is highly dependent on cloud condition. Fig. \ref{fig:mol_res} shows four cases: high-cloud, low-cloud, no-cloud, and the average of the three. The high-cloud case (orange) has the highest albedo, but line depth and density is the lowest, because the atmospheric column of absorbing gases above the cloud ($\sim$12 km) is the smallest. The effect is more notable for H$_2$O than for O$_2$, because the abundance of H$_2$O drops more than O$_2$ above the cloud. Consequently, The high-cloud case has the advantage of high reflection but the disadvantage of low spectral information content. In comparison, for the low-cloud case (red), lines are denser and deeper and the albedo is as high as the high-cloud case, so the low-cloud case represents the best case scenario for HDC observations. Lastly, the no-cloud case has the lowest albedo at $<0.1$, because the cloud-free surface is poorly reflective.

Our averaged spectrum binned down to a low spectral resolution (e.g., $R=300$) is generally consistent with Earthshine observations~\cite{Turnbull2006}. For example, the water absorption feature at 0.95 micron has an absorption depth of $\sim$50\% according to Earthshine observation (see Fig. 6 in Ref. \citenum{Turnbull2006}), consistent with our averaged spectrum (blue in Fig. \ref{fig:mol_res}). In comparison, the low-cloud case alone would have a greater continuum albedo and a deeper absorption. 


We however recognize that the cloud coverage and heights will not be known {\it a priori} for any planets to be observed. In order to cover a large parameter space, we consider all three cases for different cloud conditions: the low-cloud case, the high-cloud case, and the no-cloud case. Results for the average-cloud case and any other cases should fall within the results of the three cases, which will be discussed in subsequent sections. 

Fig. \ref{fig:mol_res} also emphasizes the need of going beyond 1 $\mu$m in order to detect biosignature gases. There are no strong lines of CH$_4$ and CO$_2$ below 1 $\mu$m. Detecting these two molecules is difficult at optical wavelengths. Even for molecules with strong lines below 1 $\mu$m, going beyond this wavelength would allow many more strong lines to be measured, thus increasing the detectability. We limit our HDC simulations below 1.8 $\mu$m, beyond which a cryogenic space mission would be required. 

\section{SIMULATIONS OF HDC OBSERVATIONS}
\label{sec:hdc}
\subsection{Methodology}
In this section, we briefly describe the procedure to simulate HDC observations and conduct data reduction. For more details in HDC simulations, please refer to two previous references\cite{Wang2016, Wang2017}. 

The planet and stellar spectra are convolved with a kernel that corresponds to a certain spectral resolution. The stellar signal is reduced by a factor that we refer to as the starlight suppression level. The starlight suppression includes the suppression from the coronagraph, wavefront control, and additional nulling boost provided by the use of a single-mode fiber\cite{Mawet2017}. 

Poisson noise is added to account for photon noise. In addition, noises incurred in data reduction are included, e.g., errors associated with removing stellar light measured by additional fibers in speckle field. One particularly important noise source, speckle chromatic noise, is also taken into account. This noise arises from wavefront control at high starlight suppression levels and prevents us from detecting biosignature gases at low spectral resolutions\cite{Wang2017}. 

We briefly discuss here how the speckle chromatic noise confuses biosignature detection at low spectral resolutions and our strategy to remove the confusion. We refer readers to \S 6.4 in Ref. \citenum{Wang2017} for more details. The speckle chromatic noise arises due to wavefront control at deep starlight suppression ($<\sim10^{-7}$). Because wavefront control is normally optimized at a certain wavelength, starlight suppression is shallower at wavelengths that deviate from the optimal wavelength. As a result, speckle would have a parabolic spectral feature that sometimes mimics an absorption band of a molecule species. We name this speckle behavior at deep starlight suppression levels as the speckle chromatic noise. The speckle chromatic noise may be difficult to distinguish from absorption features of potential biosignatures, and this noise source therefore represents a potential false positive biosignature absorption signal.

At moderately high spectral resolutions, the speckle chromatic noise is no longer a concern because simulation has shown that the noise only has low-frequency spectral features. Therefore, our strategy of removing the speckle chromatic noise is straightforward. We apply a high-pass filter to those simulated observed spectra to remove spectral features with frequencies that correspond to R$<$100, where R is spectral resolution. In this procedure, both the speckle chromatic noise and the absorption features are removed. Our HDC simulations are no longer sensitive to spectral features with R$<$100. However, the simulations are free of the influence of the speckle chromatic noise. This makes direct comparison of our technique incompatible with other works at low resolutions (R$<$100) that do not employ a high-pass filter to mitigate the speckle chromatic noise.

{{The processed spectrum is then cross correlated with a template spectrum for the molecular species of interest and at the same spectral resolution as the observation. We note that the processed spectrum contains absorption features of all molecule species and have noise appropriately incorporated and thus represents a realistic case for HDC observations}}. The resulting cross correlation function (CCF) is used for biosignature detection and to access the relative abundance of different molecules that are present in planet's atmosphere. The key parameters used in HDC simulations are given in Table \ref{tab:sun_earth}.

\begin{table}[ht]
\scriptsize
\caption{Inputs for HDC simulations for a Sun - Earth System} 
\label{tab:sun_earth}
\begin{center}       
\begin{tabular}{|l|l|l|l|l|l|} 
\hline
\multicolumn{2}{|c|}{Telescope/Instrument} &  \multicolumn{2}{|c|}{Star} & \multicolumn{2}{|c|}{Planet} \\
\hline
\rule[-1ex]{0pt}{3.5ex}  Telescope Aperture & 4 m or 12 m &   $T_\text{eff}$ & 5800 K &   Contrast & $6\times10^{-11}$ \\
\hline
\rule[-1ex]{0pt}{3.5ex}  End-to-End Throughput & 10\% &   log(g) & 4.5 &   Planet Radius & 1.0 R$_\oplus$ \\
\hline
\rule[-1ex]{0pt}{3.5ex}  Spectral Resolution & varied &   $V\sin i$ & 2.7 km/s &   $V\sin i$ & 0.5 km/s \\
\hline
\rule[-1ex]{0pt}{3.5ex}  Exposure Time & varied &   Orbital Inclination & 50 deg &   Orbital Phase & 0.25 \\
\hline
\rule[-1ex]{0pt}{3.5ex}  Wavelength & 0.5-1.8 $\mu$m &   Radial Velocity & 0.0 km/s &   Radial Velocity & 20.4 km/s \\
\hline
\rule[-1ex]{0pt}{3.5ex}  Detector Noise & 0 &  Distance & 5 pc &  Semi-major Axis & 1 AU \\
\hline
\end{tabular}
\end{center}
\end{table} 

\subsection{Detection Definition}
Each HDC simulation results in a CCF. We repeat the HDC simulation 100 times for each combination of spectral resolution and starlight suppression level. We define significance of detection in the following way. The maximum CCF value within one spectral resolution element to the planet radial velocity is recorded for each run. After 100 runs, the distribution of maximum CCF values is obtained. We then calculate 16 and 84 percentiles, half of the difference between the two values corresponds to 1-$\sigma$ of the distribution. This way of defining 1-$\sigma$ is robust against outliers. The maximum CCF value divided by 1-$\sigma$ gives detection significance. The median of detection significance values for 100 runs is recorded.

\section{RESULTS}
\label{sec:result}

In this section, we summarize the requirements for detecting certain molecular species in terms of exposure time, spectral resolution, and starlight suppression level. Listed in order of decreasing difficulty, the molecules of interest are: CH$_4$, CO$_2$, O$_2$, and H$_2$O. The order of O$_2$, and H$_2$O may switch depending on spectral resolution and starlight suppression level. Results in this section will provide the baseline requirements for the three cases with different cloud conditions.   

\subsection{Requirements for HabEx}

\subsubsection{The No-Cloud Case}
\begin{figure}
   \begin{center}
   \begin{tabular}{c}
   \includegraphics[height=15cm]{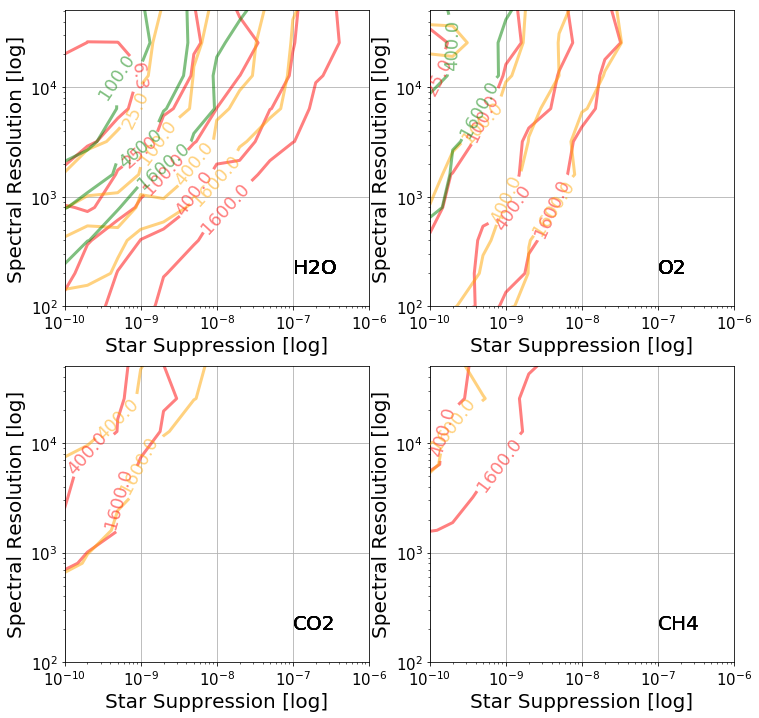}
   \end{tabular}
   \end{center}
   \caption[Exposure time contours (in hours) for 5-$\sigma$ detection] 
   { \label{fig:HabEx_avg} Exposure time contours (in hours) for 5-$\sigma$ detection. Adjacent contours differ by a factor of 4 in exposure time. The simulation is for HabEx and assumes three cases with different cloud conditions: the no-cloud case (green), the high-cloud case (orange), and the low-cloud case (red).
   }
   \end{figure}

The no-cloud case has the lowest albedo (Fig. \ref{fig:mol_res}) and thus represents the worst case scenario. CH$_4$ cannot be detected even with an exposure time of 1600 hours (as shown in green contours in Fig. \ref{fig:HabEx_avg}). Because of the low albedo, photon count is extremely low for the 4-m aperture of HabEx, there is on average 0.4 photon per pixel at the highest spectral resolution that we consider (R=51,200) for a 100-hour integration time. In addition, CH$_4$ lines are intrinsically rare and shallow at wavelengths shorter than 1.8 $\mu$m (see Fig. \ref{fig:mol_res}).  

Detection of CO$_2$ is not possible even with an exposure time of 1600 hours because of the same reason as the CH$_4$ case for an albedo smaller than 0.1. 


The minimum spectral resolution for O$_2$ detection is at R=700 and R=9,000 for 1600-hour and 400-hour exposure time and a starlight suppression level of $1\times10^{-10}$. 

High spectral resolutions are necessary to relax the requirement for starlight suppression for O$_2$ detection. For example, at R=25600, starlight suppression can be relaxed to $2\times10^{-10}$ and $1\times10^{-9}$ for 400-hour and 1600-hour exposure time.

The minimum spectral resolution for H$_2$O detection is at R$\sim$240 and R$\sim$800 for 1600-hour and 400-hour exposure time and a starlight suppression level of $1\times10^{-10}$. Starlight suppression can be relaxed to a few times $10^{-8}$ at R=51,200 for an exposure time of 1600 hours. In a comparison to the O$_2$ case, higher spectral resolutions help more for the H$_2$O case because of the abundance of water lines across the whole wavelength range. 

\subsubsection{The High-Cloud Case}
The high-cloud case has a higher albedo but line depth is much shallower than the no-cloud case. However, the high albedo outweighs the shallow line depth, making it possible to detect all four gases. 

The minimum spectral resolution for CH$_4$ detection is at R=5,000 for a 1600-hour exposure time and a starlight suppression level of $1\times10^{-10}$. It has been noted that a higher spectral resolution would help to increase detectability. However, even at R=256,000 and $1\times10^{-10}$ starlight suppression, the minimum exposure time is 850 hours.

CO$_2$ can be detected for two exposure times that we consider, 400 hours and 1600 hours, for which the minimum spectral resolution is R=8,000 and R=700 at $1\times10^{-10}$ starlight suppression, respectively. At R=51,200, the starlight suppression requirement can be relaxed to $5\times10^{-9}$ and $1\times10^{-9}$ for 1600-hour and 400-hour exposure times. 

The requirement for O$_2$ detection is less stringent than CO$_2$ detection. Spectral resolution higher than 100 will suffice for starlight suppression levels deeper than $2\times10^{-10}$ and $1\times10^{-9}$ for exposure time of 400 and 1600 hours. The requirement for starlight suppression is further relaxed toward higher spectral resolutions with increasing exposure time (see orange contours in Fig. \ref{fig:HabEx_avg}).

Because the speckle chromatic noise\cite{Wang2017} is considered in our simulation, any low frequency spectral features are removed because of a high-pass filter that we have applied in the data reduction procedure. As a result, no molecules can be detected at a spectral resolution lower than R=100.

The minimum exposure time for H$_2$O detection is 12.7 hours at at R=25,600 and C=$1\times10^{-10}$, where C denotes starlight suppression level. The requirement for starlight suppression can be relaxed by 3 orders of magnitude (compared to planet-star contrast at $\sim10^{-10}$) at high spectral resolutions (R$>$25,600) and for a long exposure time (i.e., 1600 hours). In this regime, the sensitivity boost of high dispersion coronagraphy is fully taken advantage because of sufficient photon flux and high spectral resolution.

\subsubsection{The Low-Cloud Case}
The main difference between the low-cloud case and the high-cloud case is line depth and density (see Fig. \ref{fig:mol_res}): spectral information is richer for the low-cloud case, which makes the case more amenable for spectroscopic observation. Red contours in Fig. \ref{fig:HabEx_avg} shows 5-$\sigma$ detection significance contour. Notably, the 5-$\sigma$ extends to lower spectral resolutions and shallower starlight suppression levels for H$_2$O and CH$_4$. Consequently, detection of these two molecule species is easier than the high-cloud case. This can be explained as follows: atmospheric column density for H$_2$O and CH$_4$ drops more than CO$_2$ and O$_2$, so a high cloud affects H$_2$O and CH$_4$ detectability more than CO$_2$ and O$_2$. 

For CH$_4$,  minimum exposure time is 300 hours at R=51,200 and C=$1\times10^{-10}$. For longer exposure times, the requirement for both spectral resolution and starlight suppression can be relaxed. For example, the minimum spectral resolution is at R=1,400 and R=5,000 at C=$1\times10^{-10}$ for exposure time of 1600 and 400 hours. At a spectral resolution of 51,200, starlight suppression can be relaxed to $3\times10^{-10}$ and $2\times10^{-9}$ for exposure time of 400 and 1600 hours. 

Detection of CH$_4$ in different cloud conditions exemplifies the influence of planetary spectrum on detection. The variety of possible planetary spectra may be the major uncertainty in estimating the minimum requirements for future space missions in search for biosignatures. We show here the influence of cloud condition. There are other uncertainties that may affect the mission sensitivity to biosignatures, for instance, the time evolution of Earth atmosphere. CH$_4$ concentration was much higher in Archean Earth atmosphere~\cite{Rugheimer2017}. Therefore, it is much easier to search for CH$_4$ for an Archean Earth. 

The minimum exposure time for CO$_2$ detection is 150 hours at at R=51,200 and C=$1\times10^{-10}$. The minimum spectral resolutions are R=700 and R=2,200 for exposure time of 1600 and 400 hours at a starlight suppression level of $1\times10^{-10}$. At the highest spectral resolution we consider (R=51,200), the minimum starlight suppression requirements are $7\times10^{-10}$ and $2\times10^{-9}$ for exposure time of 400 and 1600 hours. 

The requirement for O$_2$ detection is similar to that for the high-cloud case. 

The minimum exposure time for H$_2$O detection is 5 hours at at R=12,800 and C=$1\times10^{-10}$. The requirement for starlight suppression can be relaxed by 3 orders of magnitude (compared to planet-star contrast at $\sim10^{-10}$) at high spectral resolutions (R$>$25,600) and for long exposure times ($>$400 hours).  

\subsection{Requirements for LUVOIR}
\subsubsection{The No-Cloud Case}
\begin{figure}
   \begin{center}
   \begin{tabular}{c}
   \includegraphics[height=15cm]{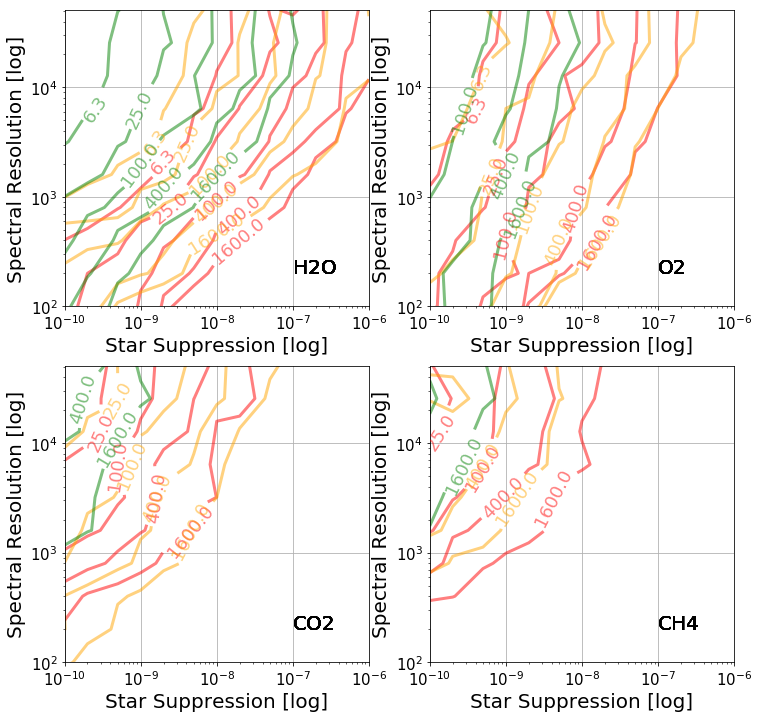}
   \end{tabular}
   \end{center}
   \caption[Exposure time contours (in hours) for 5-$\sigma$ detection] 
   { \label{fig:LUVOIR_avg} Exposure time contours (in hours) for 5-$\sigma$ detection. Adjacent contours differ by a factor of 4 in exposure time. The simulation is for LUVOIR and assumes three cases with different cloud conditions: the no-cloud case (green), the high-cloud case (orange), and the low-cloud case (red).
   }
   \end{figure}

Green contours in Fig. \ref{fig:LUVOIR_avg} shows 5-$\sigma$ detection contours at different exposure times for biosignatures for the LUVOIR no-cloud case. With a bigger aperture and thus higher photon flux, detection significance contours extend to lower spectral resolutions and shallower starlight suppression levels. Notably, the starlight suppression requirements are relaxed to $\sim10^{-8}$ and $2\times10^{-9}$ for H$_2$O and O$_2$ for an exposure time of 100 hours. We expect 100-hour as a reasonable exposure time for LUVOIR because it is a general purposed mission. CO$_2$ and CH$_4$ are not detectable given the 100-hour exposure time. The minimum exposure time for CO$_2$ and CH$_4$ detection is 183 hours and 354 hours at R=25,600 and C=$1\times10^{-10}$. At the same combination of spectral resolution and starlight suppression level, the minimum exposure time for H$_2$O and O$_2$ detection is 5.0 and 25.0 hours.   

\subsubsection{The Low-Cloud Case}
Red contours in Fig. \ref{fig:LUVOIR_avg} shows 5-$\sigma$ detection contours at different exposure times for biosignatures for the LUVOIR low-cloud case. This represents the best case scenario for biosignature search among all that we have considered in this paper. With an exposure time that shorter than 25 hours, all biosignatures can be detected at R=51,200 and C=$1\times10^{-10}$. However, we must point out that the minimum exposure time is very sensitive to what type of cloud a planet may have. The minimum exposure time usually varies by a large factor. For example, the minimum exposure time for CH$_4$ detection differs by a factor of more than 10 between the no-cloud case (354 hours) and the low-cloud case (24 hours). 

\subsubsection{The High-Cloud Case}
Orange contours in Fig. \ref{fig:LUVOIR_avg} shows 5-$\sigma$ detection contours at different exposure times for biosignatures for the LUVOIR high-cloud case. For, H$_2$O and CH$_4$, the high-cloud contours recede towards higher spectral resolutions and deeper starlight suppression (i.e., upper-left corner of the parameter space in Fig. \ref{fig:LUVOIR_avg}) when compared to the low-cloud case, indicating lower detectability of these two species in the high-cloud case. The high-cloud contours for O$_2$ are comparable between the low-cloud case and the high-cloud case, which is consistent with the HabEx result. The high-cloud contours for CO$_2$ indicate a higher detectability than the low-cloud case. However, we will show in \S \ref{sec:comp} that the detectability is in fact comparable within uncertainties in exposure time calculation. 

\section{SUMMARY AND CONCLUSION}
\label{sec:summary}
We study the detectability of four molecular species as biosignature or habitability indicators: CH$_4$, CO$_2$, O$_2$ and H$_2$O. We conduct HDC simulations to set the minimum requirements of exposure time, spectral resolution, and starlight suppression for the detection of these biosignatures. Major findings are summarized in Fig. \ref{fig:HabEx_avg}, and Fig. \ref{fig:LUVOIR_avg}. The implications are discussed in details in \S \ref{sec:result}. The results provide a baseline for mission design in order to search for biosignature gases and study the habitability of exoplanets. 

{{The results of this paper are based on HDC simulations that focus on performance at high spectral resolutions. While the simulations are compatible with low resolution results, the cross-correlation may not be the best possible way to detect planets or molecules. A more straightforward way is to conduct a conventional ADI/SDI sequence, detecting the planet, obtaining a low-resolution spectrum, and inferring the molecular presence by measuring the absorption band depth. However, we provide two arguments for using the cross-correlation technique at low spectral resolutions. First, this ensures consistency in comparing performance over a broad range of spectral resolutions. Second, speckle chromatic noise may hinder molecule detection at low spectral resolutions with conventional method, the treatment in our HDC simulation ensures the effect of the speckle chromatic noise is properly modeled and removed. }}

\subsection{Comparison to Previous Results}
\label{sec:comp}

One major difference between this work and previous works on observing Earth-like planet using low resolution spectroscopy~\cite{Robinson2016}: we consider speckle chromatic noise, which is a significant noise source at low spectral resolution (e.g., R$<$100). 

Another difference is the definition of detection significance. In this work, we adopt the local variation of CCF peaks rather than the variation estimated from other parts of CCF. In previous works, the variation was estimated using the fluctuation of first quarter and/or the forth quarter of CCF. In principle, there should be no significant difference. However, using the local variation is more representative of CCF variation around CCF peak.

\begin{figure}
   \begin{center}
   \begin{tabular}{c}
   \includegraphics[height=15cm]{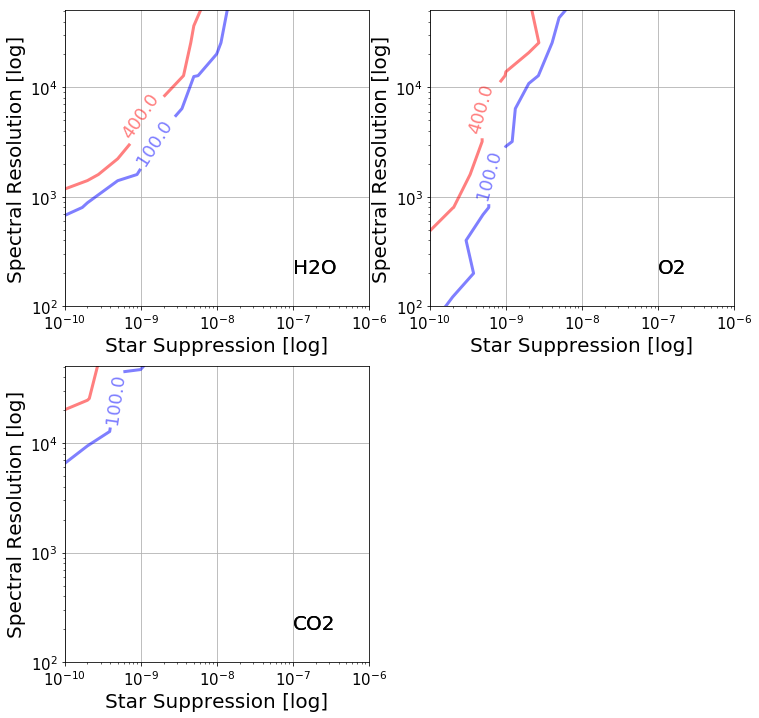}
   \end{tabular}
   \end{center}
   \caption[Comparison to previous results] 
   { \label{fig:comp} Exposure time contours (in hours) for 5-$\sigma$ detection. Blue contours are for LUVOIR and red contours are for HabEx.
   }
   \end{figure}

We compare our results to previous HDC simulations for HabEx and LUVOIR~\cite{Wang2017, Wang2017b}. Using the same input planet spectrum, i.e., the average of the no-cloud case and the low-cloud case, we compare the results for HabEx 400-hour exposure and LUVOIR 100-hour exposure (Fig. \ref{fig:comp}) to Fig. 20 and Fig. 15 in Ref. \citenum{Wang2017} to quantify the influence of difference definition of detection significance. We find that the contours in these two works track each other well with a maximum horizontal (i.e., starlight suppression level) difference by a factor of 2 at spectral resolution higher than R$\sim$2000. Part of the difference arises from statistical noise as it is evident that the contours are not smooth. The major difference between results in this paper and Fig. 20 and Fig. 15 in Ref. \citenum{Wang2017} is that we include speckle chromatic noise, which affects results at low spectral resolution significantly. Therefore, it is expected contours in Fig. \ref{fig:comp} deviate from contours in Fig. 20 and Fig. 15 in Ref. \citenum{Wang2017} at lower spectral resolutions. 

We also compare our results to Table 2 in Ref. \citenum{Wang2017b} to check the consistency of exposure time calculation for different definitions of detection significance. We can check only one HabEx case for starlight suppression level shallower than $10^{-10}$ because HDC simulations in this work do not go deeper than $10^{-10}$. At R=6,400 and C=$5\times10^{-10}$, the exposure time for O$_2$ and H$_2$O detection is 162 hours and 404 hours. These values are comparable with values in Table 2 in Ref. \citenum{Wang2017b} within a factor or 1.6. We also compare results for two LUVOIR cases (R=6400 and R=25600 at C=$1\times10^{-9}$). For R=6400 and C=$1\times10^{-9}$, the exposure time for CH$_4$, CO$_2$, O$_2$, and H$_2$O detection is $>$1600 hours, 406.2 hours, 72.1 hours, and 25.5 hours. Except for CH$_4$, exposure times for all other species are within a factor of 1.4 when compared to values in Table 2 in in Ref. \citenum{Wang2017b}. For CH$_4$, we have a lower limit for exposure time because our simulation stops at 1600 hours. For R=25,600 and C=$1\times10^{-9}$, the exposure time for CO$_2$, O$_2$, and H$_2$O detection is 196.0 hours, 39.0 hours, and 18.0 hours. These values are also within a factor of 2 to values in Ref. \citenum{Wang2017b}. 

Given the comparison with previous works, we conclude that the uncertainty of exposure time due to definition of detection significance and Poisson statistical noise is at most a factor of 2. This uncertainty is smaller than the uncertainty due to cloud condition, which can change exposure time by a factor of $\sim10$ between the low-cloud case and the no-cloud case. To account for the uncertainty of different definition of detection significance, we advise readers to use a error bar of 0.3 dex (i.e., a factor of 2) when using the minimum exposure time estimated in this paper for a given case of cloud condition. 

For the speckle chromatic noise, we use a high-pass filter to remove speckle chromatic noise. As a result, low-frequency spectral features are also removed. This results in a much longer exposure time at low spectral resolution to reach a certain detection significance. For example, the exposure time is 80 hours for R=100 and C=$1\times10^{-10}$ for O$_2$ detection. In comparison, Ref. \citenum{Robinson2016} estimated the exposure time is 200 hours for R=70 and C=$1\times10^{-10}$. The difference can be explained as follows. First, the distance of the planet in their calculation was 10 pc. Factoring the difference would reduce the exposure time from 200 hours to 50 hours.
Second, our error bar is 0.3 dex, so the actual exposure time can vary from 40 to 160 hours depending on the definition of detection significance. Therefore, the two results are consistent within uncertainty.  

\subsection{Exo-Zodiacal Flux and Thermal Backgound}
We consider in our HDC simulation only starlight suppression levels that expected to be achieved by a coronagraphic system.  However, the flux from exo-zodiacal dust may set a floor for the achievable starlight suppression level. This floor is estimated at a few times the planet-star contrast for a 4-m aperture, i.e., $\sim10^{-10}$\cite{Roberge2012, Stark2014}. LUVOIR designs are less prone to exo-zodiacal flux because of a smaller point spread function and thus smaller contamination area. This is a result of the larger aperture of LUVOIR.

Another noise source that is not accounted for in our HDC simulation is the thermal background emission from the telescope and instrument. This is not a typical concern at optical wavelengths. However, thermal background quickly becomes a severe issue at longer wavelengths. It is possible to consider a cryogenic system that cools the instrument to minimize thermal background and allows the instrument to operate at longer wavelengths. For example, mission designs exist that cool the system to 130 K without active cryogenic cooling: Euclid uses a passive cooling radiator system to cool the system to 130 K\cite{Racca2016}. To be more realistic, we will incorporate both exo-zodiacal light emission and thermal noise in future versions of HDC simulation packages.  

\acknowledgments     
 
 We would like to thank Rowan Swain for proofreading the manuscript. We thank anonymous referees for their comments and suggestions that greatly improve the paper.


\bibliography{report}   
\bibliographystyle{spiejour}   

\listoffigures
\listoftables

\end{spacing}
\end{document}